\documentclass[prb,twocolumn]{revtex4}% Physical Review B
\usepackage{epsfig}
\usepackage{graphicx}% Include figure files
\usepackage{dcolumn}% Align table columns on decimal point
\usepackage{bm}% bold math
\newcommand{\Eq}[1]{Eq.~(\ref{#1})}
\newcommand{\Eqs}[1]{Eqs.~(\ref{#1})}
\begin{document}

\title{Selective coupling of superconducting qubits via tunable
stripline cavity. }

\author{M. Wallquist}
\author{V.S. Shumeiko}%
\author{G. Wendin}
\affiliation{ Chalmers University of Technology, SE-41296
Gothenburg, Sweden.
}%

%\date{\today}
%
\begin{abstract}
We theoretically investigate selective coupling of superconducting
charge qubits mediated by a superconducting stripline cavity with a
tunable resonance frequency. The frequency control is provided by a
flux biased dc-SQUID attached to the cavity. Selective entanglement
of the qubit states is achieved by sweeping the cavity frequency
through the qubit-cavity resonances. The circuit is able to
accommodate several qubits and allows to keep the qubits at their
optimal points with respect to decoherence during the whole
operation. We derive an effective quantum Hamiltonian for the
basic, two-qubit-cavity system, and analyze appropriate circuit
parameters. We present a protocol for performing Bell inequality
measurements, and discuss a composite pulse sequence generating a
universal control-phase gate.
\end{abstract}
\maketitle

%%%%%%%%%%%%%%%%%%%%%%%%%%%%
%\section{Introduction}
%%%%%%%%%%%%%%%%%%%%%%%%%%%%%%%%%%%
%

Coherent coupling of superconducting qubits has been
experimentally demonstrated for all major qubit types
(charge\cite{Pashkin,Yamamoto}, flux\cite{Majer,terHaar,Izmalkov},
and phase\cite{Berkley,McDermott} qubits) using permanent direct
qubit-qubit coupling, capacitive or inductive. A major challenge
is to implement a tunable coupling of qubits required for any
useful gate operation. Numerous suggestions in this direction have
been discussed in recent literature together with related quantum
gate protocols (for a review see, e.g.
Ref.~\onlinecite{Handbook}).

There are two conceptually different approaches to the tunable coupling. The
first approach is to employ direct coupling schemes using Josephson junctions
in the non-resonant regime as passive controllable elements, either
capacitive,\cite{AverinBruder} or
inductive.\cite{MakhlinRev,Nori,Jonn,LyakhovBruder,Niskanen} The second
approach, which we adopt in this paper, suggests qubit coupling via a dynamic
intermediate element, e.g., $LC$-oscillator or Josephson junction, which
becomes entangled with a qubit during a two-qubit operation. In this scheme,
the entanglement is achieved by tuning the qubit and the mediator in
resonance, and then transferring the entanglement to another qubit by tuning
the mediator and the second qubit in the resonance. Such coupling method has
been first suggested\cite{Cirac} and experimentally tested\cite{Monro} for
the ion trap qubits. For superconducting qubits, qubit-oscillator
entanglement has been demonstrated experimentally for a charge qubit coupled
to a microwave stripline cavity,\cite{WallraffNature} and a flux qubit
coupled to a SQUID oscillator;\cite{Chiorescu2004,Johansson2005} the gate
protocols based on controllable qubit-oscillator coupling have been
theoretically discussed in Refs.~\onlinecite{Plastina,Blais}.

\begin{figure}[h]
\epsfxsize=6.5cm\epsffile {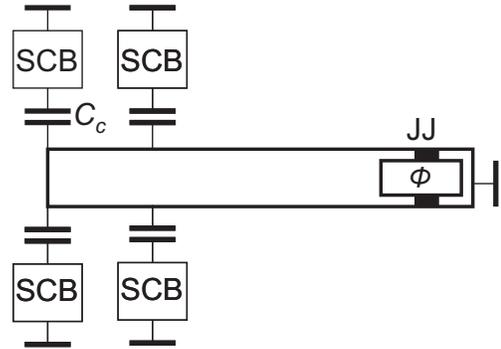} \caption{Sketch of the
device: charge qubits (single Cooper pair boxes, SCB) coupled
capacitively ($C_c$)  to a stripline cavity integrated with a
dc-SQUID formed by two large Josephson junctions (JJ); cavity
eigenfrequency is controlled by magnetic flux $\Phi$ through the
SQUID.} \label{Sketch}
\end{figure}

The experimental setup with the qubit coupling to a distributed oscillator -
stripline cavity\cite{BlaisPRA,WallraffNature} possesses potential for
scalability - several qubits can be coupled to the cavity. In this paper we
investigate the possibility to use this setup for implementation of tunable
qubit-qubit coupling and simple gate operations. Tunable qubit-cavity
coupling is achieved by varying the cavity frequency by controlling magnetic
flux through a dc-SQUID attached to the cavity (see Fig.~\ref{Sketch}). An
advantage of this method is the possibility to keep the qubits at the optimal
points with respect to decoherence during the whole two-qubit operation. The
qubits coupled to the cavity must have different frequencies, and the cavity
in the idle regime must be tuned away from resonance with all of the qubits.
Selective addressing of a particular qubit is achieved by relatively slow
passage through the resonance of a selected qubit, while other resonances are
rapidly passed. The speed of the active resonant passage should be comparable
to the qubit-cavity coupling frequency while the rapid passages should be
fast on this scale, but slow  on the scale of the cavity eigenfrequency in
order to avoid cavity excitation. This strategy requires narrow width of the
qubit-cavity resonances compared to the differences in the qubit frequencies,
determined by the available interval of the cavity frequency divided by the
number of attached qubits. This consideration simultaneously imposes a limit
on the maximum number of employed qubits. Denoting the difference in the
qubit energies, $\Delta E_J $, the coupling energy, $\kappa$, the maximum
variation of the cavity frequency, $\Delta\omega_k$, and the number of
qubits, $N$,  we summarize the above arguments with relations, $\kappa\ll
\Delta E_J$, $N\sim \hbar\Delta\omega_k/\Delta E_J$. In the off-resonance
state, the qubit-qubit coupling strength is smaller than the on-resonance
coupling by the ratio, $\kappa/(\hbar\omega_k-E_J)\ll 1$.

In the first part of the present paper, we analyze the quantum
electrical circuit consisting of superconducting stripline cavity,
dc-SQUID, and single Cooper pair box (SCB) qubits, and derive an
effective quantum Hamiltonian for this circuit, and discuss the
relevant circuit parameters.

Then, on the basis of the derived Hamiltonian, we discuss the Bell
measurement protocol and a protocol for a conditional phase gate.
We consider creating maximally entangled two-qubit states
(Bell-states) by sequentially sweeping the cavity through the
resonances with the two qubits,\cite{Plastina} and discuss the
protocol for measuring the CHSH correlation function\cite{CHSH}
for such states, which is equivalent to testing the Bell
inequality.

While considering the universal two-qubit gate, we take into
account an important feature of our system - the linearity of the
cavity, which does not allow implementation of the $\sqrt{SWAP}$
gate.\cite{Blais} We argue that the control-phase gate (CPHASE) is
a genuine two-qubit gate for our system (cf.
Ref.~\onlinecite{Plastina}). We consider a protocol for this gate,
which is much faster than the one suggested in
Ref.~\onlinecite{Plastina}, the present one
 being based on the resonant rather than
dispersive qubit-oscillator coupling. A major difficulty for
constructing such a protocol is the generation of the single- and
two-photon states in the cavity (for the cavity initialized in the
ground state); elimination of these auxiliary photon states
requires a complex pulse sequence.\cite{Schmidt-Kaler}

In this paper, we explicitly discuss the coupling of charge
qubits; however, the method of derivation of the effective quantum
Hamiltonian also applies, with minor modifications, to the flux
qubits, and the quantum protocols studied can be extended to this
type of qubit systems.

%
%%%%%%%%%%%%%%%%%%%%%%%%%%%%
\section{Cavity with variable frequency}
\label{section_cavity}

The resonant frequency of a 1D stripline cavity depends on the
boundary conditions. For example, if one end of the cavity is open
while the other is connected to the ground, the spatial
distribution of the superconducting phase along the cavity has a
maximum at the open end and a node at the grounded end. This
corresponds to the quarter-wavelength resonator, $d=\lambda/4$,
with the eigenmode wave vectors, $k_n = (\pi/d) (n+1/2)$, where
$d$ is the cavity length, the eigenmode frequencies being
$\omega_n = (\pi v/d) (n+1/2)$, where $v$ is the velocity of the
electromagnetic waves in the cavity. If, on the other hand, the
second end is disconnected from the ground, the eigenmode wave
vectors become, $k_n = (\pi/d) n$, giving the frequencies,
$\omega_n = (\pi v/d) n$. The role of the dc-SQUID attached to the
stripline cavity in Fig.~\ref{Sketch} is to vary the boundary
condition at the right end: the first case (node) corresponds to a
very large (formally infinite) Josephson energy of the SQUID,
while the second case (antinode) corresponds to the fully
suppressed Josephson energy. Thus, ideally, by changing the
biasing magnetic flux through the SQUID by half a flux quantum,
$0\leq \Phi \leq \Phi_0/2 \,({\rm mod}\,\Phi_0)$, ($\Phi_0=h/2e$),
one should be able to sweep the eigenmode frequencies within the
intervals, $(\pi v/d)n  \leq \omega_n \leq (\pi v/d)(n+1/2)$. In
practice these intervals are more narrow due to a finite maximum
and non-zero minimum Josephson energies of the SQUID.

\begin{figure}
%\begin{center}
\epsfxsize=7.5cm\epsffile{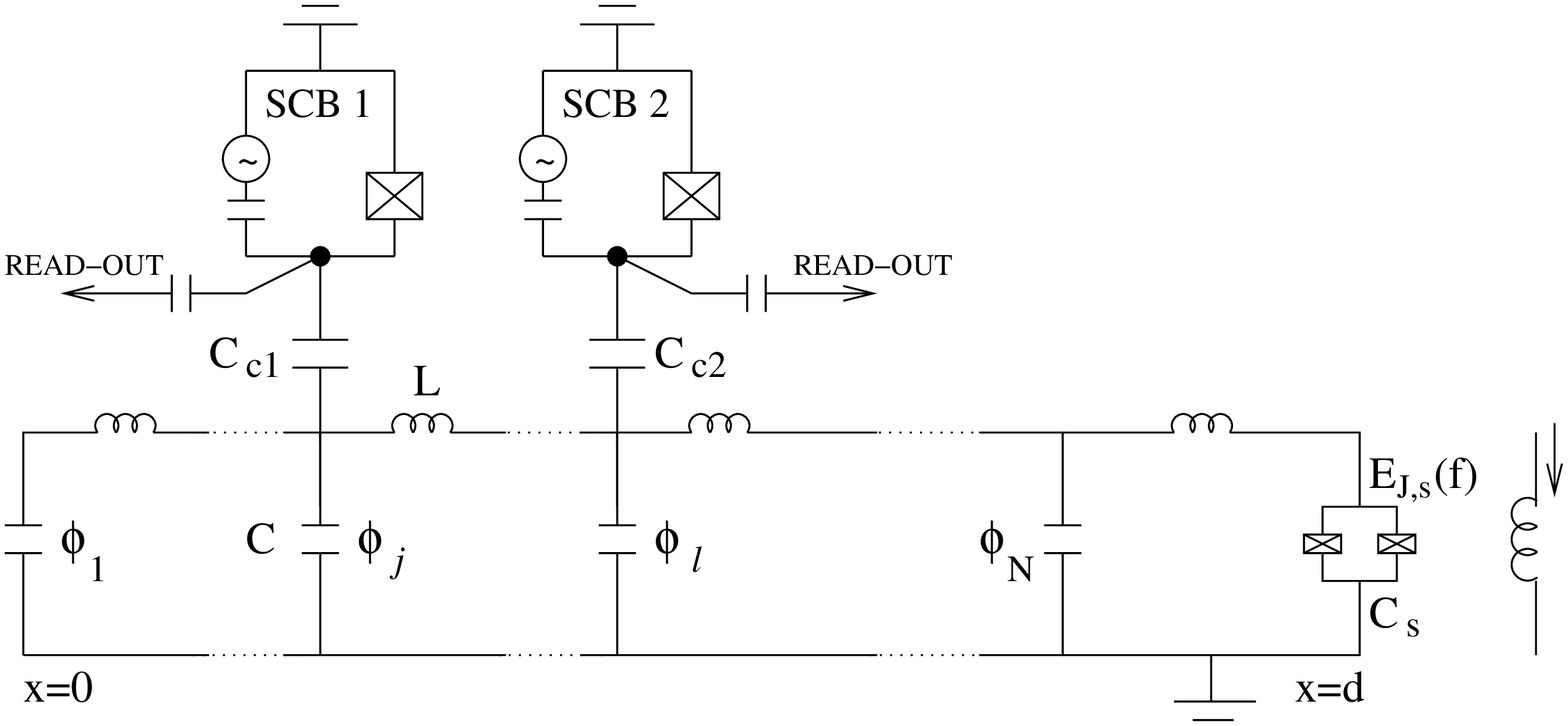}
\caption{Equivalent circuit for the device in Fig.~\ref{Sketch}:
chain of $LC$-oscillators represents the stripline cavity,
$\phi_1$ and $\phi_N$ are superconducting phase values at the ends
of the cavity, $\phi_{j}$ and $\phi_{l}$ are local phase values
where the qubits are attached; attached dc-SQUID has effective
flux-dependent Josephson energy, $E_{Js}(f)$, and capacitance
$C_s$, control line for tuning the SQUID is shown at the right;
SCB qubits are coupled to the cavity via small capacitances,
$C_{c1}$ and $C_{c2}$.} \label{2qubit_circuit}
%\end{center}
\end{figure}

For a given eigenmode, the integrated stripline\ +\ SQUID system
behaves as a lumped oscillator with variable frequency. Our goal
in this section will be to derive an effective classical
Lagrangian\cite{Yrke,Devoret} for this oscillator. To this end we
consider in Fig.~\ref{2qubit_circuit} an equivalent circuit for
the device depicted in Fig.~\ref{Sketch}. A discrete chain of
identical $LC$-oscillators, with phases $\phi_i$ across the chain
capacitors (i=1,\ldots,N), represents the stripline cavity; the dc
SQUID is directly attached at the right end of the chain, while
the superconducting Cooper pair boxes (SCB) are attached via small
coupling capacitors, $C_{c1}$ and $C_{c2}$ to the chain nodes with
local phases, $\phi_{j}$ and $\phi_{l}$ (for simplicity we
consider only two attached SCBs). The classical Lagrangian for
this circuit,
\begin{equation}
{\cal L} = {\cal L}_{SL} + {\cal L}_{squid} + \sum_{j=1,2} \left[
{\cal L}_{q,j} + {\cal L}_{coupl,j} \right], \label{L_total}
\end{equation}
consists of the stripline Lagrangian,
\begin{eqnarray}
{\cal L}_{SL} &=& \sum_{i=1}^{N-1} \left({\hbar \over 2e}
\right)^2 \left[ {C\dot{\phi}_i^2 \over 2} - {(\phi_{i+1} -
\phi_i)^2 \over 2L} \right] \nonumber\\
 &+& {\hbar^2 C  \over 2
(2e)^2} \dot{\phi}_N^2 - {\hbar^2(\phi_{s,1} - \phi_N)^2 \over 2
(2e)^2 L}, \label{L_TL_discr}
\end{eqnarray}
the SQUID Lagrangian,
\begin{equation}\label{Lsquid}
{\cal L}_{SQUID} = \sum_{i=1,2} \left[{\hbar^2 (C_s / 2) \over 2 (2e)^2}
\dot{\phi}_{s,i}^2 + E_{Js,i} \cos \phi_{s,i} \right],
\end{equation}
the Lagrangians of the SCBs,
\begin{equation}
{\cal L}_{q,j} = {\hbar^2 C_j \over 2 (2e)^2} \dot{\phi}^2_{q,j} +
{\hbar^2 C_g \over 2 (2e)^2} \left( \dot{\phi}_{q,j} + {2e \over
\hbar} V_{g,j} \right)^2 + E_{J,j} \cos \phi_{q,j} , \label{L_qj}
\end{equation}
and the capacitive SCB-stripline coupling,
\begin{equation}
{\cal L}_{coupl,j} = {\hbar^2 C_{c,j} \over 2 (2e)^2} \left(
\dot{\phi}_{j} + \dot{\phi}_{q,j} \right)^2 .\label{L_coupl_j}
\end{equation}
The SQUID junction variables are related through the flux
quantization relation, $\phi_{s,1} - \phi_{s,2} = f$, to an
externally applied magnetic flux, $\Phi = (\Phi_0 / 2\pi) f =
(\hbar / 2e) f$, threading the SQUID ring. The self inductance of
the SQUID ring is assumed to be negligibly small compared to the
Josephson inductances of the SQUID junctions. Then the SQUID can
be described as a single junction with effective capacitance,
$C_s$, and flux-dependent Josephson energy,
\begin{equation}\label{Esf}
E_{Js} (f) = [E_{Js,1}^2 + E_{Js,2}^2 + 2E_{Js,1}E_{Js,2}
\cos(f)]^{1/2}.
\end{equation}
%

%%%%%%%%%%%%%%%%%%%%%%%%%%%%%%%%%%%%%%%%%%%%%%%%%%%%%%%
\subsection{Linear approximation}

Let us assume small amplitude of the plasma oscillation in the
SQUID, $\phi_s \ll 1$, which implies the phase regime for the
SQUID, $E_{Js} (f) \gg (2e)^2 / 2C_s$, and then adopt the harmonic
oscillator approximation in \Eq{Lsquid},
\begin{equation}
{\cal L}_{SQUID} \rightarrow  {\hbar^2 C_s \over 2 (2e)^2 } \, \dot{\phi}_s^2
- { E_{Js}(f) \over 2}  \,\phi_s^2, \label{L_squid_eff}
\end{equation}
where $\phi_s = (\phi_{s,1} + \phi_{s,2})/2 + \eta (f)$, $\eta(f)$
is a constant phase shift, which can be neglected for adiabatic
flux variations. The SQUID Josephson energy $E_{Js}(f)$, \Eq{Esf},
reaches its maximum at zero magnetic flux, $E_{Js}^{max}  =
E_{Js,1} + E_{Js,2}$, while the minimum is approached at $f=\pi$:
$E_{Js}^{min}  = |E_{Js,1} - E_{Js,2}|$ with $E_{Js}^{min}>0$ due
to the SQUID asymmetry.

To proceed to a continuum description of the stripline cavity,
we introduce the distance $\Delta x$ between nodes $i$ and $i+1$,
and express the stripline Lagrangian, \Eq{L_TL_discr}, in terms of
the stripline capacitance and inductance per unit length,
$$ C_0 = C / \Delta x, \quad L_0 = L / \Delta x. $$
Let $\Delta x$ go to zero and transform the node index $i$ into
the continuous variable $x$. In the bulk of the cavity, the
equation of motion of the field is a wave equation,

\begin{equation}\label{bulk2}
\ddot{\phi} (x,t) - v^2 \phi^{\prime \prime} (x,t) = 0,
\end{equation}
where $v= 1/\sqrt{L_{0}C_{0}}$ is the wave velocity. It is
convenient to express the wave velocity through the cavity
inductance, $L_{cav}= dL_0$, and the cavity capacitance, $C_{cav}
= dC_0$,
 $$
 v = {d\over\sqrt{L_{cav}C_{cav}}}.
 $$

The boundary condition at the cavity open end ($x=0$),
\begin{equation}
 \phi^{\prime} (0,t)  = 0,
\label{RV1}
\end{equation}
requires that the only allowed solutions are of the form $\phi
(x,t) = \phi_1\sin(kvt)\cos(kx)$.
The boundary condition at the cavity right end ($x=d$) reads,
\begin{eqnarray}\label{RV2}
\phi(d,t)&=&\phi_s(t),\\
 {\hbar^2 C_s \over (2e)^2} \,\ddot{\phi} (d,t) &+& { \hbar^2 d
\over (2e)^2 L_{cav}} \,\phi^{\prime} (d,t) + E_{Js} (f) \phi
(d,t) = 0 .\nonumber
\end{eqnarray}
 A dispersion equation for
the cavity eigenmodes results from \Eq{RV2} using the bulk
solution to \Eq{bulk2}, and takes the form,
\begin{equation}
(kd) \tan (kd) = {(2e)^2 \over  \hbar^2} \,L_{cav} E_s (f) - {C_s
\over  C_{cav}} \,(kd)^2.  \label{z-rel}
\end{equation}
The solutions to this dispersion equation form an infinite set of eigenmodes
with wavelengths $\lambda = 2\pi /k$ and frequencies $\omega_k = kv $.

\begin{figure}
\begin{center}
\epsfxsize=7.5cm\epsffile {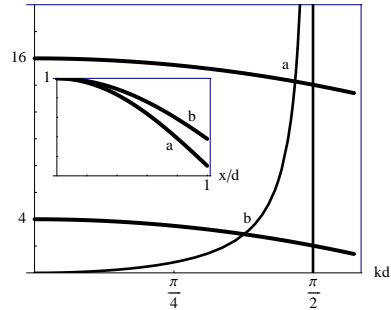}
\caption{Solution of dispersion equation (\ref{z-rel}) for first mode,
$kd\leq \pi/2$
($d \leq \lambda/4$), for large (a) and small (b) Josephson energies of
the SQUID ($(2e/\hbar)^2 L_{cav} E_s= 16$ and 4, respectively);
inset shows corresponding spatial distributions of the phase
$\phi/\phi(0)$ in the cavity.}
\label{ztanz_Lsmin}
\end{center}
\end{figure}

The solutions to Eq. (\ref{z-rel}) are illustrated in Fig.
\ref{ztanz_Lsmin}: they are given by the intersection points of
the function, $kd\tan (kd)$, with the parabola, which is almost
flat in the practically relevant limit, $C_s / C_{cav} \ll 1$. The
zeros of this function ($kd = n\pi$, $n=0,1,...$) correspond to an
open right end of the cavity (disconnected SQUID), while the
singular points ($kd = \pi/2 +\pi n$) correspond to a closed
cavity end (short circuited SQUID). These  limits of the variation
of the cavity wave eigenvectors, $n\pi\leq k_nd \leq \pi/2 +\pi
n$, can be achieved when $E_s(f)$ varies between $\infty$ and $0$;
thus ideally the frequency can be tuned between
$n\pi/\sqrt{L_{cav}C_{cav}}$ and $(\pi/2 + \pi n)/
\sqrt{L_{cav}C_{cav}}$. In practice, the available frequency range
is smaller, being limited by the value of the parameter
$[(2e)^2/(2 \hbar^2)] L_{cav} E_s(0)$, which should be chosen
large, and the minimum value of the SQUID Josephson energy,
$E_s(\pi)$, allowed by the SQUID asymmetry.

Let us return to the Lagrangian of the  stripline cavity and the
SQUID, Eqs. (\ref{L_TL_discr}) and (\ref{L_squid_eff}), and only
consider a single eigenmode, $\phi (x,t) = \phi_1 (t) \cos(kx)$.
In the continuum limit the Lagrangian will then take the form,
\begin{eqnarray}\label{Osceff}
{\cal L}_{osc} & = & \left({\hbar \over 2e} \right)^2 \int_0^d dx
\left[ {C_0\dot{\phi}_1^2 \cos^2 (k x) \over 2}
- { \phi_1^2 k^2 \sin^2 (k x)\over 2L_0} \,  \right] \nonumber \\
& + & {\hbar^2 C_s \over 2 (2e)^2 } \,\dot{\phi}_1^2 \cos^2 (k d)
- {E_s (f) \over 2} \,\phi_1^2 \cos^2 (k d).
\end{eqnarray}
After performing integration over $x$, and using the dispersion
equation (\ref{z-rel}) we arrive at the effective $LC$-oscillator
Lagrangian representing the integrated cavity+SQUID system,
\begin{equation}
{\cal L}_{osc} = {\hbar^2 C_{k} \over 2 (2e)^2 }\,
\dot{\phi}_1^2(t) - {\hbar^2 \over 2 (2e)^2 L_{k}}\, \phi_1^2(t).
\label{eff_osc_L}
\end{equation}
The oscillator is described by the effective $k$-dependent
capacitance $C_{k}$,
\begin{equation}
C_{k} = {C_{cav}\over2}  \left( 1 + {\sin (2kd)\over 2kd}\right) +
C_s \cos^2 (k d),
\end{equation}
and the effective inductance $L_{k}$,
\begin{eqnarray}
{1 \over L_{k}} & = & {(k d)^2 \over  2L_{cav}} \left(1+{\sin
(2kd)\over 2kd} + {2C_s \over C_{cav}}  \cos^2 (k d)  \right).
\label{effective_cavity}
\end{eqnarray}
The frequency of the effective oscillator, defined in the usual
way, $\omega_{k} = 1 / \sqrt{C_{k} L_{k}}$, is equal to the
frequency of the chosen cavity eigenmode as one should expect,
\begin{equation}\label{omega_k}
\omega_{k} =  {k d \over \sqrt{L_{cav}C_{cav}}}.
\end{equation}
%

%%%%%%%%%%%%%%%%%%%%%%%%%%%%%%%%%%%%%%%%%%%%%%%%
\subsection{Non-linear correction}

While the stripline cavity alone is a linear electromagnetic
system, attaching the dc-SQUID makes the integrated system
non-linear. Non-linearity will introduce a non-equidistant
correction to the quantized energy spectrum of the cavity, which
may affect the gate protocols; in particular it is harmful for the
conditional phase gate protocol considered later in the paper.
Therefore it is important to estimate non-linear effects produced
by the SQUID on the cavity.

To this end we expand the SQUID potential in the boundary
condition in \Eq{RV2}, assuming $\phi_s\ll1$, and keep a small
cubic term,
\begin{eqnarray}
&&{\hbar^2 C_s \over (2e)^2} \,\ddot{\phi} (d,t) + { \hbar^2 d
\over (2e)^2 L_{cav}} \,\phi^{\prime} (d,t)\nonumber\\
 &+& E_{Js}
(f) \left[\phi (d,t) - {1\over 6}\,\phi^3 (d,t)\right] = 0.
\label{RV2_2}
\end{eqnarray}
The cubic term will introduce the third harmonic in the cavity,
\begin{equation}\label{3harm}
\phi(x,t) = \phi_1\sin(kvt)\cos(kx) + \phi_3 \sin(3kvt)\cos(3kx),
\end{equation}
whose amplitude $\phi_3$ can be found from the boundary condition,
(\ref{RV2_2}),
\begin{eqnarray}\label{A}
\phi_3 &=& - {A_k\over 24} \,\phi_1^3,\\
 A_k &=& {\omega^2_s\cos^3(kd)\over
 [\omega^2_s - 9(kv)^2]\cos(3kd) - (3kd/L_{cav}C_s)\sin(3kd)} \,.\nonumber
\end{eqnarray}
Here we introduced the plasma frequency of the SQUID,
\begin{equation}\label{omega_s}
\omega_s^2 (f) = (2e/\hbar)^2 [E_{Js}(f)/C_s]\,.
\end{equation}
The cubic term also produces a shift of the resonance frequency
given by the corrected dispersion equation,
\begin{eqnarray}
(kd) \tan (kd) &=& {(2e)^2 \over  \hbar^2} \,L_{cav} E_s (f)
\left[1
- {1\over 8} \,\omega^2_s\cos^2(kd) \phi^2_1 \right] \nonumber\\
 &-&{C_s \over C_{cav}} \,(kd)^2\,.
\end{eqnarray}
Taking the relation (\ref{z-rel}) into account, and omitting a
small term $\sim C_s/C_k \ll 1$, we obtain the relative shift of
the frequency,
\begin{equation}\label{domega}
{\delta\omega_k\over \omega_k} = {\delta k\over k} = -
{1\over2}B_k\phi_1^2\,,\;  B_k = {(1/4)\cos^2(kd)\over
 1+ 2kd/ \sin(2kd)}\,.
\end{equation}
Such an amplitude-dependent frequency shift, on the other hand,
can be  recovered from the effective oscillator Lagrangian in
\Eq{eff_osc_L} by adding the following non-quadratic term,
\begin{equation}\label{}
\delta{\cal L}_{osc} = {\hbar^2\over 2(2e)^2 L_k}\, B_k\phi_1^4.
\end{equation}
In the quantum regime, such a term will produce a deviation from the
equidistant energy spectrum of the cavity. The magnitude of this deviation in
the first perturbative order reads,
\begin{eqnarray}\label{deltaE}
\delta E_n &=& -{\hbar^2\over 2(2e)^2 L_k}\, B_k
\langle\phi_1^4\rangle_n \nonumber\\
&=& - {6n^2+6n+3\over 4}\,B_k E_{Ck} , \,\,
\end{eqnarray}
where $E_{Ck}=(2e)^2/2C_k$ is the charging energy of the cavity,
and $n$ is the energy level number. Thus we see that the
non-linear effect is proportional to the  charging energy of the
cavity. In order to neglect the non-linear effect, this energy
must be much smaller than the energy of the qubit coupling to the
cavity (see below).

%%%%%%%%%%%%%%%%%%%%%%%%%%%%%%%%%%%%%%%%%%%%%%%%%%%%
%%%%%%%%%%%%%%%%%%%%%%%%%%%%%%%%%%%%%%%%%%%%%%%%
\section{Qubit coupling to the cavity}
\label{section_qubitcoupl}

Now we take the SCB qubits in Eq. (\ref{L_total}) into the
consideration, assuming that the coupling capacitances, $C_{c,j}$
are small enough that the perturbation of the cavity eigenmodes
due to the SCBs is negligible.

The cavity field $\phi_{j}$ at the point where SCB $j$ is coupled
is related to the effective oscillator variable $\phi_1$ by the
relation, $\phi_{j} = \phi_1 \cos (k x_{j})$, where $x_j$ is the
position of qubit $j$ along the cavity. The coupling is described
by the cross term in \Eq{L_coupl_j},
\begin{equation}
{\cal L}_{int,j} = {\hbar^2 \over (2e)^2}\, C_{c,j} \cos (k x_j)
\dot{\phi}_1 \dot{\phi}_{q,j} \,;
\end{equation}
the quadratic terms in \Eq{L_coupl_j} give small renormalization
of the qubit capacitance, $C_{\Sigma j} = C_j + C_{c,j} + C_g$,
and the oscillator capacitance.

We transform the capacitive interaction into an inductive form,
\begin{equation}
{\cal L}_{int,j} = \alpha_j E_{J,j} \phi_1 \sin \phi_{q,j},
\label{H_int_class}
\end{equation}
using the transformation\cite{Shnirman},
\begin{equation}
\phi_{q,j}  \rightarrow \phi_{q,j} + \alpha_j \phi_1,
\end{equation}
with the coupling constant,
\begin{equation}\label{alpha}
\alpha_j = C_{c,j}\cos (k x_j)/C_{\Sigma, j}, \qquad \alpha_j \ll 1.
\end{equation}
The SCB Lagrangian does not change during the transformation,
whereas the oscillator undergoes displacement,
\begin{eqnarray}
{\cal L}_{LC} &=& { \hbar^2 C_{k} \over 2 (2e)^2 }\,
\dot{\phi}_1^2 - {\hbar \over 2e} C_g (\alpha_1 V_{g,1} + \alpha_2
V_{g,2})\dot{\phi}_1 \nonumber\\
 &-& {\hbar^2  \over 2 (2e)^2
L_{k}}\, \phi_1^2, \label{L_LC_with_Vg}
\end{eqnarray}
and small renormalization of the effective capacitance, $C_{k}
\rightarrow   C_{k} - \sum_{j=1,2} C_{c,j}\cos (k x_j) [\cos (k
x_j) - \alpha_j]$.

%%%%%%%%%%%%%%%%%%%%%%%%%%%%%%%%%%%%%%%%%%%%%%%%%%%%
\subsection{Effective Hamiltonian}

We obtain the classical circuit Hamiltonian,
\begin{equation}
H = \sum_{j=1,2} \left[ H_j + H_{int,j} \right] + H_{osc},
\label{H_classical}
\end{equation}
by introducing the conjugate momenta $n = (1/\hbar)\partial L /
\partial \dot{\phi}_1$ and $n_j = (1/\hbar)\partial L / \partial
\dot{\phi}_{q,j}$. Each SCB is described by the Hamiltonian,
\begin{equation}\label{HSCB}
H_j = E_{C,j} (n_j - n_{g,j})^2 - E_{J,j}\cos\phi_{q,j},
\end{equation}
where $E_{C,j}=(2e)^2/(2C_{\Sigma, j})$ is the charging energy of
the SCB island, and $n_{g,j} = C_g V_{g,j}/(2e)$ is the
(dimensionless) charge on the island induced by the gate-voltage.

According to Eq. (\ref{L_LC_with_Vg}), the gate voltages also
induce a charge, $n_0 = \alpha_1 n_{g,1} + \alpha_2 n_{g,2}$, on
the oscillator. Because the oscillator charge is not quantized,
this induced charge does not have any physical meaning, and can be
eliminated using the gauge transformation $U^{\dagger} H_{osc} U$
with $U=\exp (-in_0\phi_1)$. The oscillator Hamiltonian then reads
\cite{timedep_term},
\begin{equation}
H_{osc} = E_{Ck} n^2 + E_{Lk} \phi_1^2,
\end{equation}
where $E_{Ck} = (2e)^2/(2C_{k})$ is the charging energy of the
oscillator and $E_{Lk} = \hbar^2/(2 (2e)^2 L_{k})$ its effective
inductive energy. The interaction term in Eq. (\ref{H_classical})
is given by the expression in \Eq{H_int_class} with the opposite
sign, $H_{int,j} = - {\cal L}_{int,j}$.

The Hamiltonian (\ref{H_classical}) is quantized by imposing the
canonical commutation relations,
$\left[\phi_{q,j},n_k\right]=i\delta_{jk}$,
$\left[\phi_1,n\right]=i$. For later convenience, the oscillator
is described in terms of the ladder operators,
$$
\phi_1 = \left( {E_{Ck} \over 4 E_{Lk}} \right)^{1/4} \left( a +
a^{\dagger} \right), \; n = i \left( {E_{Lk} \over 4 E_{Ck}}
\right)^{1/4} \left( a^{\dagger} - a \right),
$$
with $[a,a^{\dagger}]= 1$. The quantized oscillator Hamiltonian then reads,
\begin{equation}
H_{osc} = \hbar \omega_{k} \left( {1 \over 2} + a^{\dagger} a
\right),
\end{equation}
with $\omega_{k}$ given by \Eq{omega_k}.

The Coulomb blockade effect in the SCB is taken into account by
considering the periodicity of the SCB potential and imposing
$2\pi$-periodic boundary conditions on the wave function with
respect to the phase $\phi_{q,j}$. The result is charge
quantization on the island. Assuming the charge regime, $E_C\gg
E_J$ for the SCB, and keeping the system at low temperature
($k_BT\ll E_C$) and close to the charge degeneracy point
$n_g=1/2$, restricts the number of excess charges on the island to
zero or one Cooper pair. This allows us to truncate the SCB
Hilbert space to these two lowest charge states $|0\rangle$ and
$|1\rangle$ .

It is advantageous to operate at the qubit charge degeneracy
point, where the decoherence effect is minimized,\cite{Vion,Duty},
and allow only small departures, $\delta n_{g,j} = 1/ 2 -
n_{g,j}$, from this point during single qubit operations.
Considering this, we write the quantized qubit Hamiltonian in the
qubit eigenbasis at the charge degeneracy point, $[|g\rangle,
|e\rangle ] = [|0\rangle + |1\rangle, |0\rangle - |1\rangle ]$,
\begin{equation}\label{Hqb2}
H_j  =  E_{C,j} \delta n_{g,j} \sigma_{x,j} - {E_{J,j} \over 2}
\sigma_{z,j},
\end{equation}

The interaction, Eq. (\ref{H_int_class}), is proportional to
$\sin\phi_{q,j}$, which transforms into $\sigma_{y,j}$ during the
quantization procedure. It will be helpful during the discussion
of two-qubit operations to express the interaction through the
raising (lowering) qubit operators $\sigma_- |e\rangle =
|g\rangle$, $\sigma_+ |g\rangle = |e\rangle$,
$$
\sigma_{y,j} = i \left( \sigma_{+,j} -  \sigma_{-,j} \right).
$$
Thus the quantized interaction Hamiltonian reads,
\begin{equation}\label{Hint2}
H_{int,j} = i \,{\kappa_j \over 2} \left( a +  a^{\dagger} \right)
\left( \sigma_{-,j} - \sigma_{+,j} \right),
\end{equation}
where the interaction energy $\kappa_j$ is determined by the
coupling constant $\alpha_j$ in \Eq{alpha},
\begin{equation}\label{kappa}
\kappa_j = \alpha_j E_{J,j} \left( {E_{Ck} \over 4
E_{Lk}}\right)^{1/4}.
\end{equation}

Equations (\ref{Hint2}),(\ref{kappa}) were
derived for the charge limit, $E_C \gg E_J$. However, they remain
valid qualitatively also in the charge-phase regime, $E_C \sim
E_J$, which is more advantageous from the point of view of
decoherence, as is well established.\cite{Vion} In this regime, the
lowest Bloch states of the SCB Hamiltonian rather than the charge
states form the computational basis. This is fully consistent with
the quantum capacitance readout method\cite{LarsGJ1} for our
system, which realizes projective measurement on the qubit
eigenbasis. Transformation of the SCB eigenbasis from the charge
regime to the charge-phase regime with increasing ratio $E_J/ E_C$
was  analyzed in Ref. \onlinecite{NewJP}. Applying this analysis to
the present case in the Appendix we find that the qubit-cavity
coupling remains transversal acquiring the form in \Eq{HintCJapp},
\begin{equation}\label{HintCJ}
H_{int,j} =\, {a +  a^{\dagger}\over2} \left( \tilde\kappa_j
\sigma_{+,j} + \tilde\kappa^\ast_j \sigma_{-,j} \right),
\end{equation}
with the coupling constant $\tilde\kappa_j$ differing from the
coupling $\kappa_j$ of the charge regime in \Eq{kappa}  by a
complex numerical function $f(E_{J,j}/ E_{C,j})\sim 1$,
$\tilde\kappa_j=f(E_{J,j}/ E_{C,j})\kappa_j$. Such a modification
does not change qualitatively the resonant qubit-cavity dynamics
discussed in the next sections.
%
%%%%%%%%%%%%%%%%%%%%%%%%%%%%%%%%%%%%%%%%%%%%%%%%%%%%%%%%
\subsection{Constraints}

To conclude our discussion of the qubit-cavity circuit, we
summarize the imposed constraints on the circuit parameters
required for a proper functioning of the circuit.

First, we required the phase regime for the SQUID and the cavity,
implying $(2e)^2/(2C_s) = E_{Cs}\ll \hbar\omega_s \ll E_{Js}(f)$,
and $E_{Ck} \ll \hbar\omega_k\ll E_{Lk}$, respectively. The cavity
capacitance in practice greatly exceeds, by several orders, the
SQUID junction capacitances, $C_k \gg C_s$, while the cavity
inductance must be comparable to the SQUID variable inductance,
which is required by the dispersion equation (\ref{z-rel}), for
$kd\sim 1$. Thus, the SQUID plasma frequency is typically much
larger than the cavity frequency, $\omega_s\gg\omega_k$. The
latter, in its turn, must be comparable to the qubit frequencies
to provide the resonant coupling, $\omega_k \approx E_{J1}/\hbar,
E_{J2}/\hbar$.

The qubit interaction with the cavity must not be too strong in
order to provide sufficient off-resonance decoupling, $\kappa \sim
(C_c/C_\Sigma)E_J \ll |E_{J,2}- E_{J,1}|$. On the other hand, it
must exceed the variation of the level spacings in the cavity energy
spectrum, \Eq{deltaE}, $\kappa \gg E_{Ck}$, caused by non-linearity.

All these requirements can be collected in a chain of inequalities
formulating the hierarchy of relevant circuit energies,
\begin{eqnarray}\label{}
E_{Ck} \ll \kappa &\ll& |E_{J1}- E_{J2}| \leq \hbar\omega_{k} \sim
E_J,\nonumber\\
\hbar\omega_{k},\, E_{Cs} &\ll& \hbar\omega_s \ll E_{Lk} \sim
E_{Js}.
\end{eqnarray}

There is an additional requirement imposed on the lower bound of
the variation of the Josephson energy of the SQUID: the critical
current through the SQUID should be much larger than the amplitude
of the current fluctuations in the cavity. The latter is estimated
for the zero point fluctuations as $(\hbar/2e)(kd/L_{cav})
(E_{Ck}/E_{Lk})^{1/4}$, while the critical current of the SQUID is
$(2e/\hbar)E_{Js}$. Thus the requirement is equivalent, by virtue
of \Eq{z-rel}, to the inequality $\tan(kd) > (E_{Ck}/E_{Lk})^{1/4}
$.

%%%%%%%%%%%%%%%%%%%%%%%%%%%%%%%%%%%%%%%%%%%%%%%%%%%%%%%%%%
%
\section{Two-qubit operations}
\label{section_operations}
A general way of performing two-qubit operations is to
sequentially drive the cavity frequency in and out of resonance
with the qubits, i.e. $ \hbar \omega_{k} - E_{J,j} = \delta$, $|
\delta | \ll \kappa_j$. The speed should be comparable to the
scale of the qubit-cavity coupling frequency but small on the
scale of the oscillator frequency, to prevent unwanted excitation
of higher oscillator levels, $\partial_t\omega_k/\omega_k \sim
\kappa/\hbar \ll \hbar\omega_k $. During the two-qubit operations,
each qubit is parked at its charge degeneracy point, and an
appropriate difference in the qubit energies, $|E_{J,1}-E_{J,2}|
\gg \kappa$, prevents the oscillator from simultaneously
interacting with both qubits.

Consider, as an example, qubit 1 in resonance with the oscillator.
Discarding fast terms ($e^{\pm i(\omega_{k}+E_{J,1}/\hbar) t}$ and
$e^{\pm i(\omega_{k} \pm E_{J,2}/\hbar) t}$) in the interaction
picture (with $H_0 = H_1 + H_2 + H_{osc}$), which
average to zero on the time scale of the qubit-oscillator
interaction in the rotating wave approximation (RWA), the
qubit-oscillator interaction term reads,
\begin{equation}
H_{int} \rightarrow  i {\kappa_1 \over 2} \left( a^{\dagger}
\sigma_{-,1} - a \sigma_{+,1} \right).
\end{equation}
The only non-zero interaction matrix elements,
in the qubit 1 - oscillator basis, are
$\langle e,n|\sigma_+ a |g,n+1\rangle =
\langle g,n+1|\sigma_- a^{\dagger} |e,n\rangle = \sqrt{n+1}$
between
 the levels which are close to being degenerate,
$E_{n+1,g} - E_{n,e}  =  \delta$.
The (truncated) Hamiltonian of this subspace reads,
\begin{equation}\label{Hres}
H = - \ {\delta \over 2}\,\sigma_z \ + \ {\kappa_1 \over
2}\sqrt{n+1} \,\sigma_y.
\end{equation}

While discussing the entangling operations, in the following we
assume, to obtain analytical results, rectangular pulse shapes
bringing the cavity in and out of exact resonance with the qubit
($\delta =0$). The result of such an operation is given by the
unitary matrix,
\begin{equation}\label{Ures}
U(\theta) =
\left(%
\begin{array}{cc}
  \cos\theta & - \sin\theta \\
 \sin\theta  & \cos\theta\\
\end{array}%
\right), \qquad \theta = {\kappa_1\over 2}\sqrt{n+1}\;T,
\end{equation}
where $T$ is the pulse duration. A slight detuning within the
allowed interval $\delta \ll \kappa_j$ and smoother pulse shape
will not qualitatively alter the protocols.

%%%%%%%%%%%%%%%%%%%%%%%%%%%%%%%%%%%%%%%%%%%%%%%%
\subsection{Bell measurement protocol } \label{section_Bell}

Maximally entangled two-qubit states (Bell-states) are constructed
from an initial state with one excitation, say $|eg0 \rangle$, and
partly moving this excitation to the other qubit by sequentially
sweeping the oscillator through resonance with both
qubits.\cite{Plastina} The oscillator starts and ends in the
ground state.
\begin{figure}[hbt]
%\begin{center}
\epsfxsize=7.5cm\epsffile {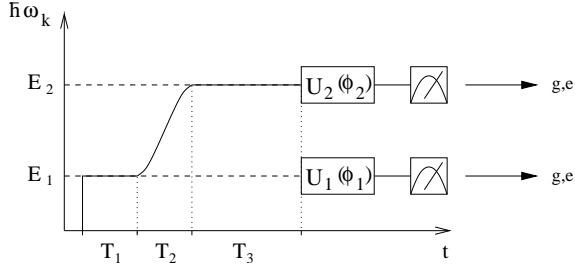}
\caption{Protocol for creating a Bell-pair: the cavity frequency
is sequentially swept through resonances with both qubits; at the
first resonance the oscillator is entangled with qubit 1, at the
next resonance the oscillator swaps its state onto qubit 2 and
ends up in the ground state. A Bell measurement is performed by
applying Rabi pulses to non-interacting qubits, and projecting on
the qubit eigenbasis, $\{|g\rangle,|e\rangle\}$, by measuring
quantum capacitance. } \label{fig_prot_Bell}
%\end{center}
\end{figure}
The first pulse brings the cavity to resonance with the excited
qubit 1, $\hbar \omega_{k} = E_1$ (see Fig.~\ref{fig_prot_Bell})
during a time  satisfying the relation, $\kappa_1 T_1 = \hbar \pi
/2$ ($\pi/2$-pulse). Then the cavity, which has become entangled
with qubit 1,
$$
|eg0 \rangle \rightarrow {1 \over \sqrt{2}} \left( |eg0 \rangle +
|gg1 \rangle \right),
$$
is driven towards resonance with qubit 2. The accumulated phase
during the free evolution ($\hbar \omega_{k} \neq E_1, E_2$ for a
time $T_2$) is $\varphi_2$,
\begin{equation}
{1 \over \sqrt{2}} \left( |eg0 \rangle + e^{i\varphi_2}|gg1 \rangle
\right), \qquad \varphi_2= \int_0^{T_2}\left[{E_1 \over \hbar} -
\omega_k(t)\right]\,dt .
\end{equation}
After the oscillator has reached resonance with qubit 2, $\hbar
\omega_{k} = E_2$,  it stays in resonance during the time, $\kappa_2
T_3 = \hbar \pi $ ($\pi$-pulse). The system evolves to the state,
$$
{1 \over \sqrt{2}} \left( |eg0 \rangle - e^{i(E_1 - E_2)T_3/\hbar}
e^{i\varphi_2}|ge0 \rangle \right).
$$
Choosing the time $T_2$ of free evolution such that the accumulated
phase, $\varphi_2$, satisfies the equation, $ \exp[i(E_1 - E_2)T_3/\hbar
+ i\varphi_2] = \mp 1$, the two-qubit system ends up in one of the
Bell states,
\begin{equation}
|\Psi_{\pm} \rangle = {1 \over \sqrt{2}} \left( |eg \rangle \pm |ge
\rangle \right) |0\rangle. \label{bellstate}
\end{equation}
Note that the cavity has returned to its ground state.

To perform a  test of the non-classical statistical properties of
the Bell state, \Eq{bellstate}, it is convenient to implement a
protocol similar to the one of Ref.~\onlinecite{Rowe} for
measuring the CHSH inequality \cite{CHSH}. The protocol consists
of two independent rotations of the uncoupled qubits by means of
applying Rabi pulses, and then measuring the quantum capacitances
of both SCB \cite{LarsGJ1}. The latter realizes a projective
measurement on the qubit eigenstates, $|g_j\rangle$,
$|e_j\rangle$.

The resonant $\pi/2$-pulse applied to the gate of the $j$-th
qubit,
\begin{equation}\label{pulse}
\delta n_{g,j} (t) = \delta n_{g,j}^0 \cos (E_{J,j} \,t ),
\end{equation}
during the time $T_a$ such that $T_a= \hbar \pi /(2E_C \delta
n_{g,j}^0)$ produces a unitary transformation,
\begin{equation}\label{RabiU}
U_j (\phi_j) = \left(
\begin{array}{cc}
1 &  -i e^{-i\phi_j}\\
-i e^{i\phi_j}  & 1
\end{array}
\right), \quad \phi_{j,a} = E_{J,j}T_a/\hbar,
\end{equation}
By applying such pulses to both qubits, one gets the state,
\begin{equation}\label{Psiphi}
\Psi_{\pm}(\phi_1, \phi_2) = U_1(\phi_1)U_2(\phi_2)\Psi_\pm.
\end{equation}
After the measurement, the qubits will be found either in
similar states (both qubits in the ground, $|gg\rangle$, or excited,
$|ee\rangle$, states), or in different states ($|ge\rangle$ or
$|eg\rangle$). The corresponding correlation function reads,
\begin{equation}\label{q}
q_\pm(\phi_1, \phi_2)= \langle \Psi_{\pm}(\phi_1,
\phi_2)|\sigma_{z,1}\sigma_{z,2}| \Psi_{\pm}(\phi_1, \phi_2)\rangle.
\end{equation}

By repeating the procedure with a $\pi/2$-pulse of slightly
different duration, $T_b$, with $T_b-T_a \sim \hbar/E_{J,j}\ll
T_a$, we obtain the four correlation functions,
\begin{equation}\label{qs}
q(\phi_{1a}, \phi_{2a}), \;  q(\phi_{1b}, \phi_{2a}), \;
q(\phi_{1a}, \phi_{2b}), \; q(\phi_{1b}, \phi_{2b}).
\end{equation}
According to the analysis in Ref.~\onlinecite{CHSH}, the quantity,
\begin{equation}\label{B}
B = |q(\phi_{1a}, \phi_{2a})- q(\phi_{1b}, \phi_{2a})| +
|q(\phi_{1a}, \phi_{2b}) + q(\phi_{1b}, \phi_{2b})|,
\end{equation}
has an upper bound for classical statistics, $B\leq 2$. For
the state in \Eq{bellstate}, however, we have,
\begin{equation}\label{qBell}
q_\pm (\phi_1,\phi_2) = \pm \cos (\phi_1 -
\phi_2),
\end{equation}
and the upper bound for $B$ becomes $B\leq 2\sqrt 2$, the equality
being achieved for,
\begin{equation}\label{}
\phi_{1a} = - \phi_{2a} = {3\pi\over 8}, \quad \phi_{1b} =
-\phi_{2b} = -{\pi\over 8}.
\end{equation}

%%%%%%%%%%%%%%%%%%%%%%%%%%%%%%%%%%%%%%%%%%%%%%%%
\subsection{Two-qubit control-phase gate} \label{section_cphase}

\begin{figure}[hbt]
%\begin{center}
\epsfxsize=6.5cm\epsffile {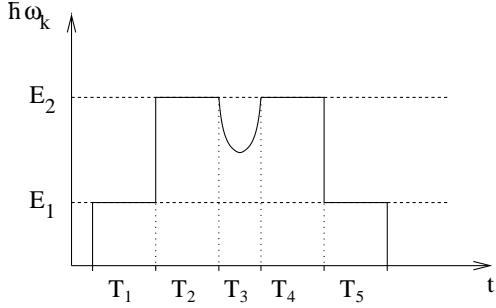}
\caption{Pulse sequence producing (trivial) diagonal gate: during
time $T_1$, qubit 1 swaps its state onto the oscillator, then the
oscillator interacts with qubit 2 before swapping its state back
onto qubit 1; free evolution during time $T_3$ is added to
annihilate two-photon state in the cavity. }
\label{fig_prot_naive}
%\end{center}
\end{figure}

In this section we modify the Bell state construction to
implementing a control-phase (CPHASE) two-qubit gate. This gate
has the diagonal form: $|\alpha \beta 0 \rangle \rightarrow
\exp(i\phi_{\alpha \beta}) |\alpha \beta 0 \rangle$ ($\phi_{00} =
\phi_{01} = \phi_{10} = 0$, $\phi_{11} = \pi$), and it is
equivalent to the CNOT gate (up to local rotations). To generate
such a diagonal gate, we adopt the following strategy: first tune
the oscillator through resonance with both qubits performing
$\pi$-pulse swaps in every step, and then reverse the sequence, as
shown in figure \ref{fig_prot_naive}. With an even number of swaps
at every level, clearly the resulting gate will be diagonal.

Such a strategy would indeed produce the CPHASE gate provided the
qubit-cavity coupling-constant in \Eq{Hres} does not depend on
the number of photons. Indeed, applying the rectangular
$\pi$-pulse to pass the first resonance, $\kappa_1T_1=\pi\hbar$,
and then the $\pi$-pulse for the second resonance,
$\kappa_2T_2=\pi\hbar$, and then reversing the pulse sequence,
$\kappa_2T_4=\pi\hbar$, and $\kappa_1T_5=\pi\hbar$, see Fig.~
\ref{fig_prot_naive}, we induce the following transitions,
\begin{eqnarray}\label{stateevolution}
\begin{array}{rrrrrrrrr}
|gg0\rangle &\stackrel{1}\rightarrow & |gg0\rangle &
\stackrel{2}\rightarrow & |gg0\rangle &\stackrel{4} \rightarrow &
|gg0\rangle &\stackrel{5} \rightarrow & + |gg0\rangle \\
|ge0\rangle &\rightarrow & |ge0\rangle &\rightarrow & |gg1\rangle &\rightarrow
&
-|ge0\rangle & \rightarrow & -|ge0\rangle \\
|eg0\rangle &\rightarrow & |gg1\rangle &\rightarrow & -|ge0\rangle
&\rightarrow & -|gg1\rangle &\rightarrow & +|eg0\rangle \\
|ee0\rangle &\rightarrow & |ge1\rangle &\rightarrow & |gg2\rangle
& \rightarrow & -|ge1\rangle & \rightarrow & +|ee0\rangle,\\
\end{array}
\end{eqnarray}
which generate the CPHASE gate. The problem is, however, that it
is not possible to make the swaps  for {\em all} the states at the
second resonance, e.g., $|ge0\rangle \rightarrow |gg1\rangle$, and
$|ge1\rangle \rightarrow |gg2\rangle$, with the {\em same} pulse
because of different values of the coupling constants, $\kappa_2$,
and $\kappa_2\sqrt2$, respectively. Thus the $\pi$-pulse for the
first transition will necessarily produce a state superposition
for the second transition and vice versa. There is a possibility
to annihilate this superposition by inserting an interference loop
in the pulse sequence, see Fig.~\ref{fig_prot_naive}, namely, by
departing from the resonance for a while to accumulate an
appropriate phase shift during the free evolution; the required
time, $T_3$ for such an excursion is given by equation,
\begin{equation}
\theta_3= \int_0^{T_3}\left[{E_2 \over \hbar} -
\omega_k(t)\right]\,dt = \pi \;({\rm mod}\,2\pi) .
\label{teta_3_cphase}
\end{equation}
However, it is easy to check that such a protocol will generate a
non-entangling, trivial gate.

\begin{figure}[hbt]
\epsfxsize=8.0cm\epsffile {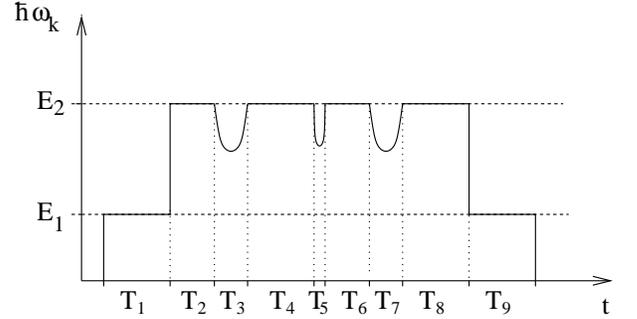}
\caption{ Correct pulse sequence for performing a control-phase
gate: time intervals $T_1$, $T_4$, $T_8$ and $T_9$ are
single-photon $\pi$-pulses, whereas $T_2$ and $T_6$ are two-photon
$\pi$-pulses; free evolution during times $T_3, T_5$ and $T_7$ is
added to annihilate excited photon states, and create non-trivial
phase shift. } \label{fig_prot_SK}
\end{figure}

It turns out that by adding two more swap segments with
interference loops to the previous protocol  it is possible to
obtain a sufficient amount of free parameters to annihilate the
state superpositions, and to obtain an entangling
gate\cite{Schmidt-Kaler}. The pulse sequence is shown in
Fig.~\ref{fig_prot_SK}, and it consists, at the second resonance,
of the two $\pi$-pulses producing swaps to the single-photon
states, $ \kappa_2 T_4 =  \kappa_2 T_8 = \hbar\pi$, as well as the
two $\pi$-pulses producing swaps to the two-photon states, $\sqrt
2 \kappa_2 T_2 =  \sqrt 2\kappa_2 T_6 = \hbar\pi$. The first and
the third interference loops are included in the pulse sequence to
annihilate the state superpositions, the corresponding phase
shifts satisfying the relations,
\begin{eqnarray}\label{theta357}
\theta_3 & - &\theta_5 =\pi, \quad \theta_5-\theta_7 =\pi \; ({\rm mod}\,2\pi)
, \\
\displaystyle\theta_n & = &\int_0^{T_n}\left[{E_2 \over \hbar} -
\omega_k(t)\right]\,dt \nonumber .
\end{eqnarray}
The middle loop is required for generating a non-trivial gate.
%%%%%%%%%%%%%%%%%%%%%%%%%%%%%%%%%%%%%%%%%%%%%%%%%%%%%%%%%
Specifically, the pulse sequence ($T_2$ - $T_8$) produces on the states
$|ge0\rangle$ and $|gg1\rangle$ the following gate operation,
$$
U ({\pi \over 2}) \ S_7 \ U({\pi \over 2\sqrt{2}}) \ S_5 \ U ({\pi \over 2}) \
S_3 \ U({\pi \over 2\sqrt{2}}) =
e^{-i\theta_5} \left(
\begin{array}{cc}
1 & 0 \\
0 & e^{-i\theta_5} \end{array} \right),
$$
where $S_n = \textrm{diag}[e^{-i\theta_n},1]$. For the levels
$|ge1\rangle$ and $|gg2\rangle$, the sequence produces the gate
operation,
$$
U ({\pi \over \sqrt{2}}) S_7 U ({\pi \over 2}) S_5 U({\pi \over \sqrt{2}})
S_3 U ({\pi \over 2}) =
e^{-i\theta_5} \left(
\begin{array}{cc}
1 & 0 \\
0 & e^{-i\theta_5} \end{array} \right).
$$

Incorporating the modified gate operation into the sequence,
Eq.~(\ref{stateevolution}),
we end up with the state evolution corresponding to a diagonal gate operation,
\begin{eqnarray}\label{}
\begin{array}{rrrrrrr}
|gg0\rangle &\stackrel{1}\rightarrow & |gg0\rangle &
\stackrel{2 - 8}\rightarrow &
|gg0\rangle &\stackrel{9} \rightarrow & + |gg0\rangle \\
|ge0\rangle &\rightarrow & |ge0\rangle &\rightarrow & e^{-i\theta_5}
|ge0\rangle
& \rightarrow & e^{-i\theta_5}|ge0\rangle \\
|eg0\rangle &\rightarrow & |gg1\rangle &\rightarrow & e^{-i 2\theta_5}
|gg1\rangle
 &\rightarrow & - e^{-i2 \theta_5}|eg0\rangle \\
|ee0\rangle &\rightarrow & |ge1\rangle &\rightarrow & e^{-i\theta_5}
|ge1\rangle
& \rightarrow & - e^{-i\theta_5}|ee0\rangle .\\
\end{array}
\end{eqnarray}
The overall protocol produces the universal CPHASE gate (up to a
common phase and single-qubit rotations),
\begin{equation}\label{final_gate}
\begin{array}{r}
|gg0\rangle \\
|ge0\rangle \\
|eg0\rangle \\
|ee0\rangle 
\end{array}
\
\rightarrow
\ 
\left[
\begin{array}{cccc}
1 & & &  \\
 & 1 & &  \\
 & & e^{-i\theta_5} &  \\
 & & & e^{i\theta_5}
\end{array}
\right]
\begin{array}{r}
|gg0\rangle \\
|ge0\rangle \\
|eg0\rangle \\
|ee0\rangle 
\end{array}.
\end{equation}

\begin{figure}[hbt]
%\begin{center}
\epsfxsize=8cm\epsffile {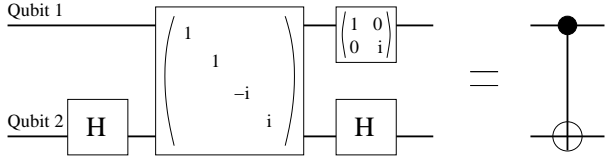}
\caption{Gate circuit for constructing a CNOT gate using the
control-phase gate: a z-axis rotation is applied to qubit 1,
and Hadamard gates H are applied to the second qubit.}
\label{fig_cnot-circuit}
%\end{center}
\end{figure}

For a specific choice of the
angle, $\theta_5 = \pi/2 ({\rm mod}\,2\pi)$, the gate in
\Eq{final_gate} can be further transformed to a CNOT gate using
Hadamard rotations on qubit 2 as shown in
Fig.~\ref{fig_cnot-circuit}.

%
%%%%%%%%%%%%%%%%%%%%%%%%%%%%%%%%%%%%%%%%%%%%%%%%%%%%%%%%%
%%%%%%%%%%%%%%%%%%%%%%%%%%%%%%%%%%%%%%%%%%%%%%%%%%%%%%%%%%
\section{Discussion}
\label{section_disc}

The protocol for the CPHASE gate considered in the previous
section is much faster than the one suggested for a similar
qubit-oscillator system and based on the dispersive
qubit-oscillator coupling\cite{Plastina} (by the ratio of the
coupling frequency to the deviation from the exact resonance
(detuning)). On
the other hand, the CPHASE gate protocol discussed in this paper
is more complex and relatively slower than the protocol for direct
qubit-qubit $zz$-coupling considered in
Refs.~\onlinecite{Jonn,NewJP}: the duration of the gate operation
in the latter case is $h/8$ in the units of inverse coupling
energy, while it is $2.7h$ for the protocol presented in Fig.
\ref{fig_prot_SK}. This illustrates the advantage of longitudinal,
$zz$ coupling (in the qubit eigenbasis), which is achieved for the
charge qubits biased at the charge degeneracy point by
current-current coupling.\cite{Jonn,NewJP} More common for charge
qubits is the capacitive coupling, however there the situation is
different: this coupling has $xx$ symmetry at the charge degeneracy
point, and
because of inevitable difference in the qubit frequencies, the
gate operation takes much longer time, prolonged by the ratio between
the qubits frequency asymmetry and the coupling frequency. Recent
suggestions to employ dynamic control methods to effectively bring
the qubits into resonance\cite{Rigetti,Bertet} can speed up the
gate operation. For these protocols, the gate duration is $\sim h$
in units of direct coupling energy, which is longer than in the
case of $zz$ coupling, but somewhat shorter than in our case.
However, the protocol considered in this paper might be made
faster by using pulse shaping.

A specific feature of the present protocol is that there is always
at least one qubit staying off-resonance: one qubit stays
off-resonance during manipulations with the other qubit, and also
the second qubit undergoes off-resonance excursions during the
manipulation. During these periods, free evolution of the qubits is
assumed, however, a finite off-resonant qubit-cavity coupling
violates this assumption and eventually negatively affects the
fidelity of the gate. Therefore it is particularly important for
this protocol to provide a weak off-resonance interaction. The
latter is perturbatively estimated as $\sim \kappa^2/E_J$, which is
by factor, $\sim C_c/C_\Sigma$, smaller than the resonant
interaction. In practice this factor would not exceed 1/100
(coupling frequency $\sim 80$ MHz for qubit frequency $>5$ GHz),
which would provide the fidelity of the gate not worse than one
percent.

To estimate a maximum number of qubits the circuit is able to
accommodate, we assume reduced coupling strength, $\kappa/h \sim
10$ MHz (which would still allow up to 10 two-qubit operations
during optimistic 1 $\mu$sec decoherence time). Assuming the same gate
fidelity, 1\%, and given the fact that the maximum qubit frequency
is in the range of $E_J/h \sim 10$ GHz, we find that the number of
qubits can not be more than ten qubits. Thus we conclude that the
resonant method of selective qubit addressing considered in the
present paper allows to implement a small quantum processor
suitable for testing the simplest quantum algorithms, although it
has limited potential for larger scale applications.

\begin{acknowledgments}
Acknowledgment - We thank Chris Wilson, Per Delsing, Martin
Sandberg, G\"oran Johansson, and Lars Tornberg
 for helpful discussions.
Support by
the European Union under contracts IST-3-015708-IP
EuroSQIP and IST-SQUBIT-2, by SSF-NANODEV, and by the Swedish Research
Council is gratefully acknowledged.

\end{acknowledgments}

%%%%%%%%%%%%%%%%%%%%%%%%%%%%%%%%%%%%%%
%%%%%%%%%%%%%%%%%%%%%%%%%%%%%%%%%%%%%%%%%%%%%%%%%
\appendix
\section{Charge-phase regime}

In this appendix we derive \Eq{HintCJ} for the qubit-cavity interaction
in the charge-phase regime.

The starting point is the SCB Hamiltonian in \Eq{HSCB} at the
degeneracy point, $n_g = 1/2$,  written in the charge basis,
$|n\rangle$,
\begin{eqnarray}
H &=& \sum_{n=-\infty}^{\infty}\left[E_C
(n-1/2)^2|n\rangle\langle n| \right. \nonumber\\
  &-& \left.(E_J/2)\left(|n+1\rangle\langle
n|+|n-1\rangle\langle n|\right)\right],
\end{eqnarray}
We split the complete set of the charge eigenstates, $-\infty< n <
\infty$, in the positive and negative charge subsets labelled with
$\sigma = \uparrow,\downarrow$, and $m=\ldots, 2,1$ such that
\begin{eqnarray}
m = n, & n>0,\nonumber\\
m = 1-n, \qquad& n \leq 0.
\end{eqnarray}
In the basis $|m,\sigma\rangle$, the Hamiltonian $H$ acquires the
form,
\begin{eqnarray}\label{H1}
H &=& \left[
\begin{array}{cc}
A & B \\
 B & A
\end{array}\right],
\end{eqnarray}
where $A$ is tridiagonal matrix,
\begin{equation}
A = {E_C\over 4}\left[
\begin{array}{cccc}
%\ddots & \ddots & & \\
%\ddots & 16 & -2E_J/E_C & \\
\ddots & \ddots & \ddots & \\
 & -p & 9 & -p \\
 & & -p & 1
\end{array}\right], \;\;p= {2E_J\over E_C},
\end{equation}
and matrix $B$ contains only a single element,
\begin{equation}
 B = -{E_J\over 2}\left[
\begin{array}{ccc}
\ddots & & \vdots\\
 & 0 & 0\\
\ldots & 0 & 1
\end{array}\right].
\end{equation}
Similarly, the interaction Hamiltonian, corresponding to
\Eqs{H_int_class}, (\ref{alpha}), (\ref{kappa}),
\begin{equation}\label{}
H_{int}= {i\kappa\over 2}(a + a^\dagger)\sum^\infty_{n=\infty}
\left(|n+1\rangle\langle n|-|n-1\rangle\langle n|\right),
\end{equation}
takes the form in the $(m,\sigma)$-representation,
\begin{eqnarray}\label{}
H_{int} &=& (a + a^\dagger)\left[
\begin{array}{cc}
C & iD \\
 -iD & -C
\end{array}\right],
\end{eqnarray}
where
\begin{equation}
C = {\kappa\over 2}\left[
\begin{array}{cccc}
\ddots & \ddots & \ddots& \\
%\ddots & 0 & 1 & \\
& -i & 0 & i \\
 & & -i & 0
\end{array}\right], \quad
D = {\kappa\over 2}\left[
\begin{array}{ccc}
\ddots & & \vdots\\
 & 0 & 0\\
\ldots & 0 & 1
\end{array}\right].
\end{equation}

 A Hadamard rotation, H, in $\sigma$-space,
\begin{equation}\label{Hadamard}
{\textrm H}=\frac{1}{\sqrt 2}\left(\sigma_z+\sigma_x\right),
\end{equation}
takes the basis $|m\uparrow\rangle$, $|m\downarrow\rangle$ to \\
$|m\pm\rangle = (1/\sqrt{2}) (|m\uparrow\rangle \pm
|m\downarrow\rangle)$, and transforms the matrices in \Eqs{H1} and
(\ref{}), to the form,
\begin{equation}\label{diagH1}
H = A {\mathbf 1} + B\sigma_z,\quad
 H_{int} = (a + a^\dagger)(C\sigma_x + D\sigma_y) .
\end{equation}
The eigenstates of the SCB Hamiltonian, the Bloch states, which
consist of superpositions of many charge states,  can be found by
independent unitary rotations in the $\sigma = +$ and $\sigma = -$
subspaces due to block-diagonal form of the matrix $H$ in
\Eq{diagH1}.

The qubits states are chosen as the two lowest energy Bloch states.
The corresponding energy eigenvalues occupy the bottom right
corners of the diagonalized matrices, $[U^\dagger(A+B)U]_{11}$ and
$[V^\dagger(A-B)V]_{11}$, where $U(p)$ and $V(p)$ are appropriate
unitary matrices. Indeed, this is obviously true for the charge
regime ($p\ll 1$), when these matrix elements correspond to
superpositions of the $n=0$ and $n=1$ charge states. When the
Josephson energy increases, which corresponds to increasing
potential energy in the Bloch Hamiltonian ($p\sim 1$), the Bloch
energies change without crossing \cite{BenderOrszag}, and therefore
the bottom right corner matrix elements remain to be the lowest
energy eigenvalues.

In the qubit subspace, the interaction Hamiltonian in \Eq{diagH1}
takes the form,
\begin{equation}\label{Hint1}
H_{int} = (a + a^\dagger)\left[
\begin{array}{cc}
0 & [U^\dagger(C-iD)V]_{11} \\
\left[V^\dagger(C+iD)U\right]_{11} & 0
 %_{11} & 0
\end{array}\right].
\end{equation}
Denoting the matrix element of the Hermitian matrix in \Eq{Hint1}
as $\tilde\kappa/2 = f(p) \kappa/2$, where $f(p)$ is a complex
numerical coefficient, we arrive at the qubit-cavity interaction on
the form,
\begin{equation}\label{HintCJapp}
H_{int} =\, {a +  a^{\dagger}\over2} \left( \tilde\kappa \sigma_{+}
+ \tilde\kappa^\ast \sigma_{-} \right).
\end{equation}
%
%%%%%%%%%%%%%%%%%%%%%%%%%%%%%%%%%%%%%%%%%%%%%%%%%%%%%%%%%%%%%
%%%%%%%%%%%%%%%%%%%%%%%%%%%%%%%%%%%%%%%%%%%%%%%%%%%%%%%%%%%%

%%%%%%%%%%%%%%%%%%%%%%%%%%%%%%%%%%%%%%%%%%%%%%%%%%%

\begin{thebibliography}{999}
\bibitem{timedep_term}
During gate voltage operations, the additional term
$iU^{\dagger}\partial_t U$ becomes relevant, shifting the
oscillator to $E_{Lk}(\phi_1 + \dot{n}_0 / (2E_{Lk}))^2$. The
shift is small compared to the gate voltage speed $\dot{n}_{g,i}$
because $\alpha \ll 1$.
\end{thebibliography}
\end{document}